\documentclass[prb,twocolumn,showpacs,showkeys,superscriptaddress]{revtex4-1}

\usepackage[utf8]{inputenc}
\usepackage[english]{babel}
\usepackage{setspace}
\usepackage{graphicx}
\usepackage{subfigure}
\usepackage{amssymb}
\usepackage{amsmath}
\usepackage{amscd}
\usepackage{rotating}
\usepackage{color}
\usepackage{enumerate}
\usepackage{multirow}

\usepackage{hyperref}
\hypersetup{
    colorlinks=true,
}

\usepackage{amsthm}

\begin{document}

\renewcommand{\v}[1]{\ensuremath{\mathbf{#1}}} 
\newcommand{\gv}[1]{\ensuremath{\mbox{\boldmath$ #1 $}}} 
\newcommand{\md}[1]{\mathrm{d}#1\,} 
\renewcommand{\d}[2]{\frac{d #1}{d #2}} 
\newcommand{\dd}[2]{\frac{d^2 #1}{d #2^2}} 
\newcommand{\pd}[2]{\frac{\partial #1}{\partial #2}} 
\newcommand{\pdd}[2]{\frac{\partial^2 #1}{\partial #2^2}} 
\newcommand{\pdc}[3]{\left( \frac{\partial #1}{\partial #2} \right)_{#3}} 
\newcommand{\ket}[1]{\left| #1 \right>} 
\newcommand{\bra}[1]{\left< #1 \right|} 
\newcommand{\braket}[2]{\mathopen{}\mathclose\bgroup\left< #1 \vphantom{#2} \aftergroup\egroup\right| \mathopen{}\mathclose\bgroup\left. #2 \vphantom{#1}  \aftergroup\egroup\right>} 
\newcommand{\brakett}[2]{\left< #1 \vphantom{#2} \right| \left. #2 \vphantom{#1}  \right>} 
\newcommand{\matrixel}[3]{\left< #1 \vphantom{#2#3} \right| #2 \left| #3 \vphantom{#1#2} \right>} 
\newcommand{\Var}[1]{\mathrm{Var}[ #1 ]}

\newcommand{\eqnref}[1]{(\ref{#1})}

\newcommand{\bela}[1]{[\emph{\color{red} BB: #1}]}
\newcommand{\mt}[1]{[\emph{\color{blue} MT: #1}]}
\newcommand{\dolfim}[1]{[\emph{\color{magenta} MD: #1}]}

\title{Pair Correlations in Doped Hubbard Ladders}

\author{Michele Dolfi}
\affiliation{Theoretische Physik, ETH Zurich, 8093 Zurich, Switzerland}

\author{Bela Bauer}
\affiliation{Microsoft Research, Station Q, University of California, Santa Barbara, CA 93106, USA}

\author{Sebastian Keller}
\affiliation{Laboratorium f\"ur Physikalische Chemie, ETH Zurich, 8093 Zurich, Switzerland}

\author{Matthias Troyer}
\affiliation{Theoretische Physik, ETH Zurich, 8093 Zurich, Switzerland}

\date{\today}

\begin{abstract}
Hubbard ladders are an important stepping stone to the physics of the two-dimensional Hubbard model. While many of their properties are accessible
to numerical and analytical techniques, the question of whether weakly hole-doped Hubbard ladders are dominated by superconducting or charge-density-wave
correlations has so far eluded a definitive answer. In particular, previous numerical simulations of Hubbard ladders have seen a much faster decay
of superconducting correlations than expected based on analytical arguments.
We revisit this question using a state-of-the-art implementation of the density matrix renormalization group algorithm that allows us to
simulate larger system sizes with higher accuracy than before.
Performing careful extrapolations of the results, we obtain improved estimates for the Luttinger liquid parameter and the correlation
functions at long distances. Our results confirm that, as suggested by analytical considerations, superconducting correlations become
dominant in the limit of very small doping.
\end{abstract}

\pacs{71.27.+a, 74.20.Rp, 74.72.Gh, 02.70.-c}



\maketitle

\section{Introduction}

The question of whether electrons in two dimensions can exhibit superconductivity mediated by repulsive interactions, which is
motivated by the discovery of high-temperature superconductors, has become
one of the central questions of condensed matter theory. However,
the numerical study of even the simplest models, such as the Hubbard or $t$-$J$ model, are made difficult by a multitude of competing
low-energy phases that these models exhibit. This is particularly the case in the regime of weak doping away from half filling, which is most
relevant for the phase diagram of cuprate superconductors. Numerical efforts reviewed in Ref.~\onlinecite{scalapino2007} as well as more recent results in Refs.~\onlinecite{corboz2014,leblanc2015} have shown a close competition of striped antiferromagnetic phases, $d$-wave superconducting phases, and other more exotic phases such as a pseudogap phase where hole quasi-particles play the role of the mobile carriers.

Faced with these challenges, quasi-one-dimensional systems such as ladders have appeared as an easier starting point
to investigate the properties of these models, as they are amenable to a broader range of numerical and analytical
methods. These approaches view the system as essentially one-dimensional with additional degrees of freedom that allow the
two-dimensional characteristics to emerge. Crucially, this has allowed treatment using the density matrix renormalization group (DMRG),~\cite{white1992,white1992-1} which allows accurate simulations of extended quasi-one-dimensional systems
and has successfully illuminated many properties of ladder systems.\cite{schollwoeck2005}

Numerical work on $t$-$J$ and Hubbard ladders\cite{dagotto1992,tsunetsugu1994,troyer1996} as well as
analytical work on the weak-interaction limit by Balents and Fisher~\cite{balents1996} has shown that in a wide parameter
regime, weakly doped ladders fall into the Luther-Emery universality class,\cite{luther1974,haldane1980} which has a gapped spin mode and a single
gapless charge mode. This phase is a possible precursor phase to two different ordered phases in the two-dimensional
limit, a superconducting phase (SC) and a charge-density wave (CDW) phase. To distinguish these phases,
it is crucial to compute whether the ladder system is dominated by density-density or superconducting
correlations. Within the Luther-Emery universality class, these both decay with a power-law whose exponents
are determined by a single dimensionless parameter $K_\rho$.

Previous DMRG calculations of the correlation functions\cite{noack1996,noack1997} have observed power-law decay
of the correlation functions, but have found a surprisingly fast decay of the superconducting correlations inconsistent
with dominant superconducting correlations.
This decay was found to be inconsistent with calculations in the weak-doping limit,\cite{schulz1999,siller2001} and
also violates certain identities of the Luther-Emery universality class.
Here, we revisit the calculation of these exponents with a focus on extracting the correct behavior of the pair correlation
function. We exploit the advances in DMRG methods, in particular on the correct extrapolation of physical quantities, and increases
in computational power since the work of Refs.~\onlinecite{noack1996,noack1997}.
Using a high-performance DMRG code,\cite{dolfi2014} we are able to target longer systems with much improved accuracy to obtain reliable correlation exponents for the two-leg Hubbard ladder, which settle the disagreement between the numerical calculations and the theoretical expectations. To achieve this, we carefully analyze the effects of the finite system size and the DMRG truncation on correlation functions. 

The paper is organized as follows. In Section~\ref{sec:model} we present the Hubbard model. In Sections~\ref{sec:method} we briefly introduce the DMRG method and how correlation observables are extrapolated to the thermodynamic limit. Sections~\ref{sec:friedel-oscillations} and~\ref{sec:correlations} are devoted to the discussion of our results: first the finite size analysis of density oscillations, then the comparison of the Luttinger liquid exponent with the pair and density correlation functions. In Section~\ref{sec:conclusions} we present our conclusions.

\begin{figure}[t!]
\centering
\includegraphics[width=\columnwidth]{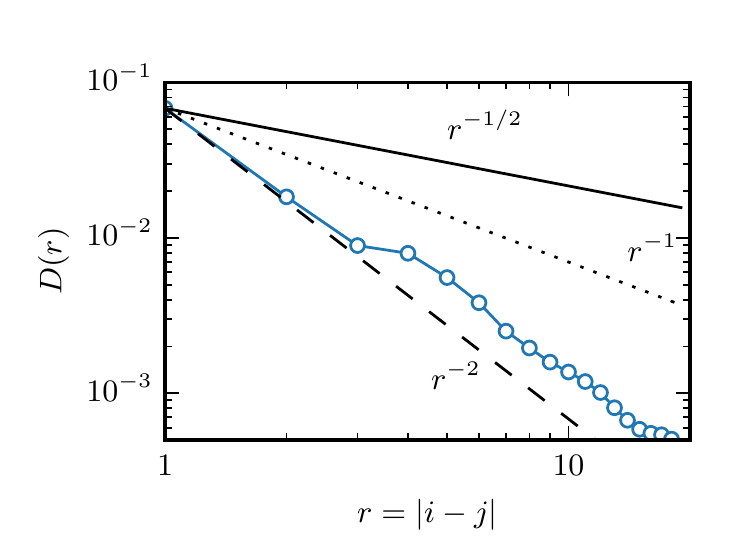}
\caption{(color online) Spatial decay of the pair correlation function (blue markers) for a $2\times 32$ Hubbard model with $U/t=8$ and average filling $n=0.875$ similar to the results obtained in Ref.~\onlinecite{noack1997}. The solid, dotted and dashed lines are reference power-law decays with exponents $\mu=-1/2, -1, -2$, respectively. In this paper we estimate an exponent $\mu\approx -1$ (see exponents in Table~\ref{tab:Krho-Friedel}).}
\label{fig:pairfield-short}
\end{figure}

\section{\label{sec:model}The Hubbard ladder}
We consider the Hubbard model on a two-leg ladder described by the Hamiltonian
\begin{align}
\label{eq:hamiltonian}
\hat H =& -t \sum_{i, \lambda, \sigma} \left[ \hat c^\dagger_{(i,\lambda), \sigma} \hat c_{(i+1,\lambda),\sigma} + \text{H.c.} \right] \notag \\
&-t_\perp \sum_{i, \sigma} \left[ \hat c^\dagger_{(i,1),\sigma} \hat c_{(i,2),\sigma} + \text{H.c.} \right] \\
& + U \sum_{i,\lambda} \hat n_{(i,\lambda),\uparrow} \hat n_{(i, \lambda),\downarrow} \notag ,
\end{align}
where the index $i$ runs along two coupled chains of length $L$ and $\lambda=1,2$ identifies the two chains respectively. The operator $\hat c^\dagger_{(i,\lambda),\sigma}$ creates a fermion at site $i$ on chain $\lambda$ with spin $\sigma \in \{\uparrow, \downarrow\}$ and $\hat n_{(i,\lambda),\sigma} = \hat c^\dagger_{(i,\lambda),\sigma} \hat c_{(i,\lambda),\sigma}$.

The phases of this model can conveniently be labeled by the number of gapless spin and charge modes, with up to two gapless modes
possible in each sector. Much attention has been focused on the phase with one gapless charge mode, but a gap in the spin sector
(labeled C1S0 in Ref.~\onlinecite{balents1996}) for its relevance for both superconducting (SC) and charge-density wave (CDW) phases
in the two-dimensional limit. This phase is found in a wide parameter range
of repulsive $U$, $t_\perp < 2t$, and hole-doping with filling $n < 1$ (in units where one fermion per site corresponds to $n=1$).
In this paper we will focus on the isotropic hopping case $t_\perp=t$ with interaction $U/t=8$, for which the spin gap has previously been reported\cite{noack1996} to show a maximum. We investigate different values of the average
filling $n$ while keeping the total magnetization fixed at zero.

We define the local rung density operator as $\hat n_i=\sum_{\lambda,\sigma} \hat n_{(i,\lambda),\sigma}$ and its expectation value as
\begin{equation}
n_i=\sum_{\lambda,\sigma} \langle \hat n_{(i,\lambda),\sigma} \rangle.
\end{equation}
Its density correlation function is
\begin{equation}
\label{eq:dens_corr}
N(i,j) = \langle \hat n_i\, \hat n_j \rangle - \langle \hat n_i \rangle \langle \hat n_j \rangle,
\end{equation}
and the $d$-wave pair correlation function takes the form
\begin{equation}
\label{eq:pair_corr}
D(i,j) = \langle \hat \Delta_i^\dagger \hat \Delta_j \rangle,
\end{equation}
where $\hat \Delta_i^\dagger = \hat c^\dagger_{(i,1),\uparrow}\, \hat c^\dagger_{(i,2),\downarrow} - \hat c^\dagger_{(i,1),\downarrow}\, \hat c^\dagger_{(i,2),\uparrow}$ creates a singlet on rung $i$.

In the Luther-Emery phase, the spatial decay of the density-density correlation function $N(r)$ and the pair correlation function $D(r)$ at large distance $r$ are dominated by a power-law parametrized by the non-universal parameter $K_\rho$:
\begin{align}
\label{eq:dens-exponents} N(r) \propto r^{-\nu} \quad & \text{with} \quad \nu=K_\rho, \\
\label{eq:pairfield-exponents} D(r) \propto r^{-\mu} \quad & \text{with} \quad \mu=1/K_\rho.
\end{align}
Because of the relation $\nu \cdot \mu=1$, one has that for $K_\rho > 1$ the system is dominated by the $d$-wave pair correlations, whereas for $K_\rho < 1$ one observes dominant charge density wave correlations. The Luttinger liquid parameter $K_\rho$ must in general be determined numerically.
In the limit $n \to 1$ and in the strong-coupling limit of the $t$-$J$ model, one can construct an effective bosonic model for hole pairs in the Hubbard model,\cite{schulz1999,siller2001} which yields a universal power-law decay of the pair correlation function $D(r) \propto 1 / \sqrt{r}$, and thus $K_\rho = 2$.
Previous DMRG calculations, whose results we reproduce in Fig.~\ref{fig:pairfield-short}, found the decay of
the pair-correlation function to be much faster than $D(r) \propto 1 / \sqrt{r}$ for
weak doping, and a comparison of the decay of pair- and density-density correlations showed a violation of the identity $\nu \cdot \mu = 1$.

\section{\label{sec:method}Simulation method}

\subsection{The density matrix renormalization group algorithm}
We tackle the model using an implementation of the density matrix renormalization group (DMRG) method\cite{white1992,white1992-1} in a formalism
of matrix-product states (MPS)\cite{fannes1992,ostlund1995} available as part of the ALPS project.\cite{bauer2011-alps,dolfi2014} For recent reviews of these
methods, see Refs.~\onlinecite{schollwock2011,schollwoeck2005}.


The DMRG method can be understood as a variational optimization over Matrix Product State (MPS) wavefunctions, which are a class
of one-dimensional ansatz states that can be systematically refined by increasing the so-called bond dimension $M$. In a standard approach,
the variational optimization proceeds by iteratively improving the wavefunction on pairs of sites. In each optimization step, a truncation occurs,
and the sum of the discarded components of the wave function, called truncated weight $\varepsilon$, is stored for later evaluations (see Section~\ref{sec:extrapolations}).

While the MPS ansatz is exact for $M\rightarrow\infty$, a finite value of $M$ restricts the amount of entanglement that can be captured by the wave function. It has been shown~\cite{hastings2007,schuch2008-1,verstraete2006-2} that this is an efficient representation of the ground states of one-dimensional, gapped, local Hamiltonians. For gapless systems with a dynamical critical exponent of $z=1$, one finds that only a polynomially growing bond dimension~\cite{vidal2003a,latorre2004} is generally required to accurately describe local properties. When coupling chains to ladders, $M$ has to increase exponentially with the width of the ladders, as the entanglement entropy grows linearly with the width.

\begin{figure}[t]
\centering
\includegraphics[width=\columnwidth]{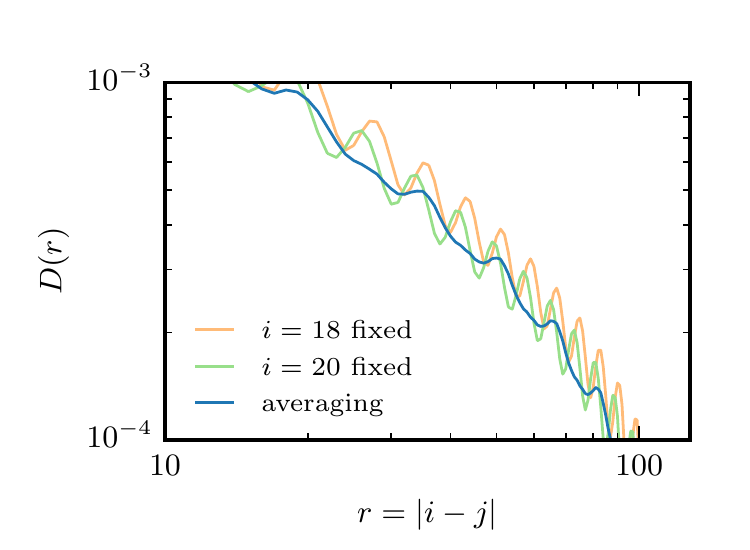}
\caption{(color online) Spatial decay of the pair correlation function $D(r)$ on ladders with $L=128$ and $n=0.875$ as a function of distance $r=|i-j|$ for several choices of $i$: fixing $i=18$ (orange line) and $i=20$ (green line), and averaging 11 pairs $(i,j)$ at distance $r$ around the middle according to Eq.~\eqref{eq:averaging} (blue line).}
\label{fig:pairfield-starts}
\end{figure}

\begin{figure}[t]
\centering
\includegraphics[width=\columnwidth]{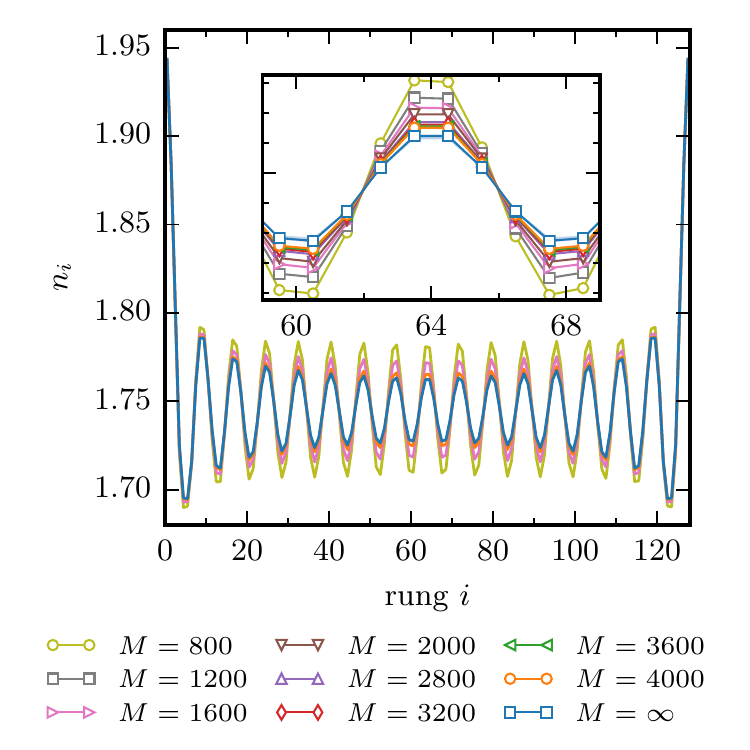}
\caption{(color online) Local density profile for ladders with $L=128$ and $n=0.875$. for several bond dimensions $M$, the inset showing details in the center of the system. The uncertainty due to systematic errors in the extrapolated curve $M=\infty$ is shown as a shaded region around the estimate; its size is often smaller than the symbols. }
\label{fig:density-bonddim}
\end{figure}

\begin{figure}[t]
\centering
\includegraphics[width=\columnwidth]{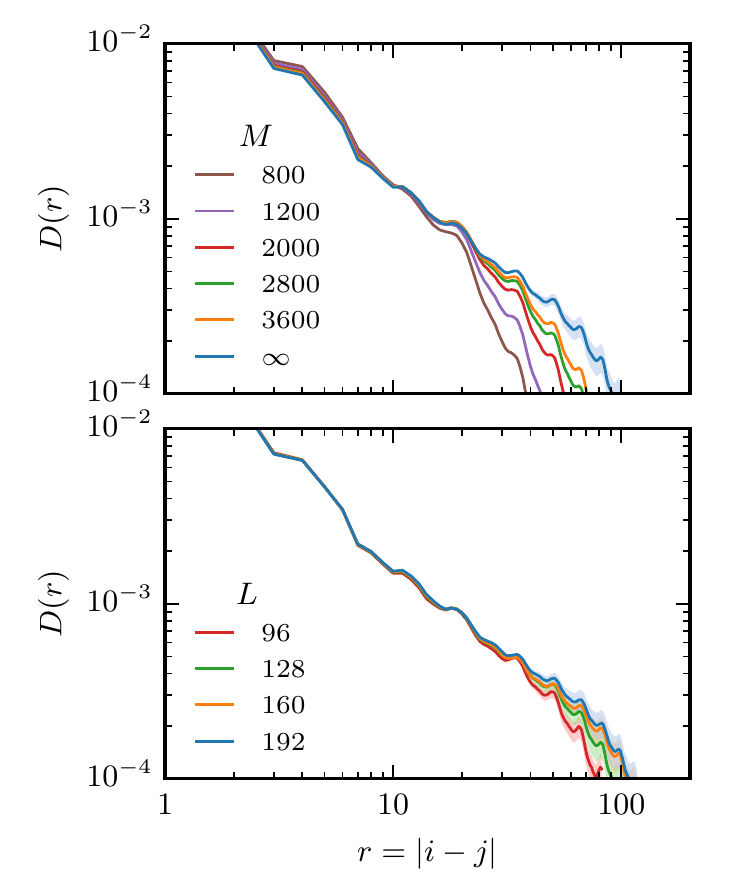}
\caption{(color online) Spatial decay of the pair correlation function $D(r)$ for ladders with $n=0.875$ and several bond dimensions $M$ and system sizes $L$. Top panel: Results for $L=128$. Bottom panel: Results are extrapolated to $M=\infty$. The shaded region around the curve show the confidence range of the systematic error.}
\label{fig:pairfield-bonddim}
\end{figure}

DMRG is most efficiently performed with open boundary conditions, which causes complications since this breaks translational invariance. Local quantities (for example density or magnetization) will differ from site to site and correlation functions depend not only on the distance $r=|i-j|$ between two sites but on both sites $i$ and $j$. The latter effect can be observed in Fig.~\ref{fig:pairfield-starts}, where we plot $D(r)$ at various distances between one fixed site $i$ and an other site $j=i+r$. 
To reduce boundary effects for a correlation function $C(r)$, we evaluate the value at distance $r$ by averaging over 11 pairs $r=|i-j|$ around the middle of the system
\begin{equation}
\label{eq:averaging}
C(r) = \frac{1}{11} \sum_{s=-5}^5 C\left(\left\lfloor\frac{L - r}{2}\right\rfloor + s, \left\lfloor\frac{L + r}{2}\right\rfloor + s\right).
\end{equation}
The solid dark line in Fig.~\ref{fig:pairfield-starts} shows the impact of averaging for reducing the oscillations induced by the boundaries.

In the simulations shown here, we use a two-site update algorithm and generally perform between 20 and 30 sweeps until energy and local observables converge for a given bond dimension $M$. For all model parameters, we perform independent simulations with several bond dimensions up to a maximum of $M=4800$.

\subsection{Finite size and finite entanglement scaling}

To obtain reliable long-range correlation functions, DMRG results need to be extrapolated both in system size $L$ and bond dimension $M$. In Figs.~\ref{fig:density-bonddim} and \ref{fig:pairfield-bonddim} we show the local density and the pair correlation functions for various values of $M$ and $L$. It is apparent in Fig.~\ref{fig:pairfield-bonddim} that calculations with insufficient system size or bond dimension may lead to underestimating the strength of the correlations.

To understand the interplay of bond dimension and system size, it is crucial to note that a matrix-product state always
exhibits exponentially decaying correlations at large enough distances,\cite{fannes1992} but may reproduce power-law
decay at short distances. The scale on which the correlations cross over from power-law to exponential behavior, $\xi_M$, is dictated by
the bond dimension $M$. At the same time, the finite size of the system will introduce some length scale $\xi_L$ on which
boundary and finite-size effects become significant, and correlations are no longer representative of the thermodynamic limit.
In interpreting results obtained with DMRG for finite $M$ and $L$, it is important to distinguish different regimes depending on
whether deviations from the asymptotic behavior are dominated by $\xi_M$ or by $\xi_L$.
Following Ref.~\onlinecite{pirvu2012}, the regime of small bond dimension $M$, where $\xi_M \ll \xi_L$, is referred to as finite entanglement
scaling (FES) regime, in which the system does not feel the presence of the boundaries because the correlation length induced by
the finite bond dimension is short ranged compared to the system size.
In the other limit, where $\xi_M$ exhausts $\xi_L$ and correlations can in principle span the whole system,
the finite size scaling (FSS) regime is reached. In an intermediate regime, where $\xi_M$ and $\xi_L$ are comparable,
a two-parameter scaling may be necessary.

To illustrate these regimes, the top panel of Fig.~\ref{fig:pairfield-bonddim} shows $D(r)$ for several bond dimensions $M$ at $L=128$.
We can clearly distinguish how correlations are cut off at a certain length scale $\xi_M$ depending on the bond dimension; as the bond
dimension $M$ is increased, we can consider the correlations converged for the given system size over an increasing range of distances.
The bottom panel shows results that have been extrapolated in $M$ for different system sizes $L$, and thus suffer only from finite-size
corrections.
By comparing the two panels, we see that for the range of system sizes considered here, $\xi_L \approx 50$ and thus the results
in the upper panel suffer primarily from finite-entanglement corrections for $r < 50$. For the bond dimensions we can attain in
practice, corrections due to $\xi_M$ set in at shorter distances, as seen in the upper panel.


In light of these considerations, we avoid having to perform a two-parameter scaling and focus mostly on the more
relevant corrections due to finite entanglement. We thus perform careful extrapolations in the bond dimension $M$
for a fixed, given system size, achieving the finite-size scaling limit $\xi_M \gg \xi_L$,
and then compare results obtained for different system sizes to assess the reliability. Most of the results
in the remainder of this paper are obtained for $L=128$.

\subsection{Extrapolation to infinite bond dimension}
\label{sec:extrapolations}

When extrapolating observables to infinite bond dimension $M=\infty$ one choice is to extrapolate in $1/M$. However, a more controlled extrapolation may be possible by extrapolating in the variance of the energy
\begin{equation}
\label{eq:variance}
\Var{\hat H} = \langle \hat H^2 \rangle - \langle \hat H \rangle^2,
\end{equation}
which vanishes for an eigenstate of $\hat H$. Similarly, the truncated weight $\varepsilon$ will vanish when the bond dimension $M$ is large enough to faithfully represent the wave function. For a more reliable data analysis we compare extrapolations in both $1/M$, truncated weight $\varepsilon$ and $\Var{\hat H}$. Unless noted otherwise, we will plot the average of the three extrapolations together with a confidence interval given by the minimum and maximum extrapolated values as an estimate of the systematic error. When further analysis is performed, {\it e.g.} for the determination of $K_\rho$, the analysis is performed for both extrapolations to obtain error estimates on the results.

\subsubsection{Extrapolating the ground state energy}

\begin{figure}[t]
\centering
\includegraphics[width=\columnwidth]{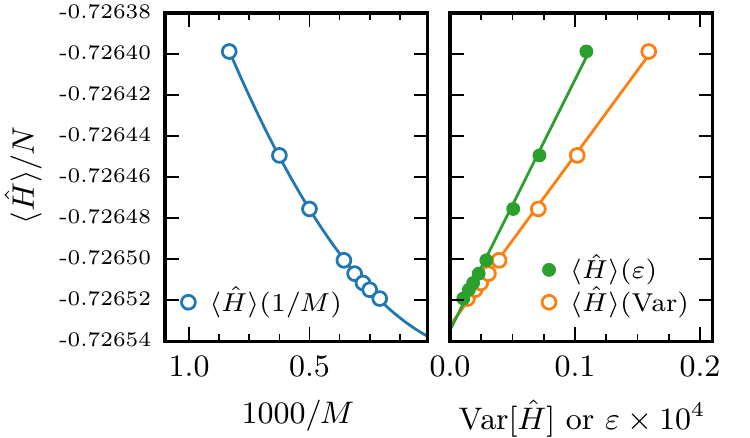}
\caption{(color online) Extrapolation of the ground state energy density as a function of the inverse bond dimension $1/M$ (left panel), the truncated weight $\varepsilon$ (right panel, full circles) and as a function of the energy variance $\Var{\hat H}$ (right panel, empty circles). Results are for $L=128$ and $n=0.875$.}
\label{fig:energy-extrap}
\end{figure}

For the ground state energy, the deviation of $\langle \hat H \rangle$ from the ground state energy $E_0$ is known\footnote{For example this has also been used in variational Monte Carlo techniques~\cite{sorella2001}.} to depend linearly on $\Var{\hat H}$, which provides for very accurate extrapolations.
To demonstrate this dependence we write the state $\ket{\psi} = \ket{\psi_0} + \ket{\delta}$ obtained by DMRG as the sum of the true ground state $\ket{\psi_0}$ with energy $E_0$ and an error term $\ket{\delta}$ with $\langle \delta | \delta \rangle = \delta^2$. Both $\ket{\psi}$ and $\ket{\psi_0}$ are supposed to be normalized. The energy of this state then is\
(with $\tilde{\delta} = \braket{\psi_0}{\delta}  + \braket{\delta}{\psi_0}$)
\begin{equation}
\label{eq:dmrg-energy}
\langle \hat H \rangle = \langle\psi|\hat H|\psi\rangle = E_0(1+\tilde{\delta}) + O(\delta^2).
\end{equation}
Similarly, the expectation value of $\hat H^2$ is
\begin{equation}
\label{eq:dmrg-energy2}
\langle \hat H^2 \rangle = E_0^2(1+\tilde{\delta})+ O(\delta^2).
\end{equation}
Combining these results we obtain for the energy variance~\eqref{eq:variance}:
\begin{equation}
\langle \hat H^2 \rangle - \langle \hat H \rangle^2 = E_0^2 \tilde{\delta} + O(\delta^2),
\end{equation}
which can be used to derive a linear dependence of the expectation value of the ground state energy on the variance:
\begin{equation}
\langle \hat H \rangle = E_0 + a \Var{\hat H} + O(\delta^2),
\end{equation}
where $a$ is a non-universal pre-factor independent of the DMRG error $\delta$.
This linear dependence can be seen in Fig.~\ref{fig:energy-extrap}, and provides a more reliable extrapolation than by extrapolating na\"ively in $1/M$.

\subsubsection{Extrapolating other observables}
\begin{figure}[t!]
\centering
\includegraphics[width=\columnwidth]{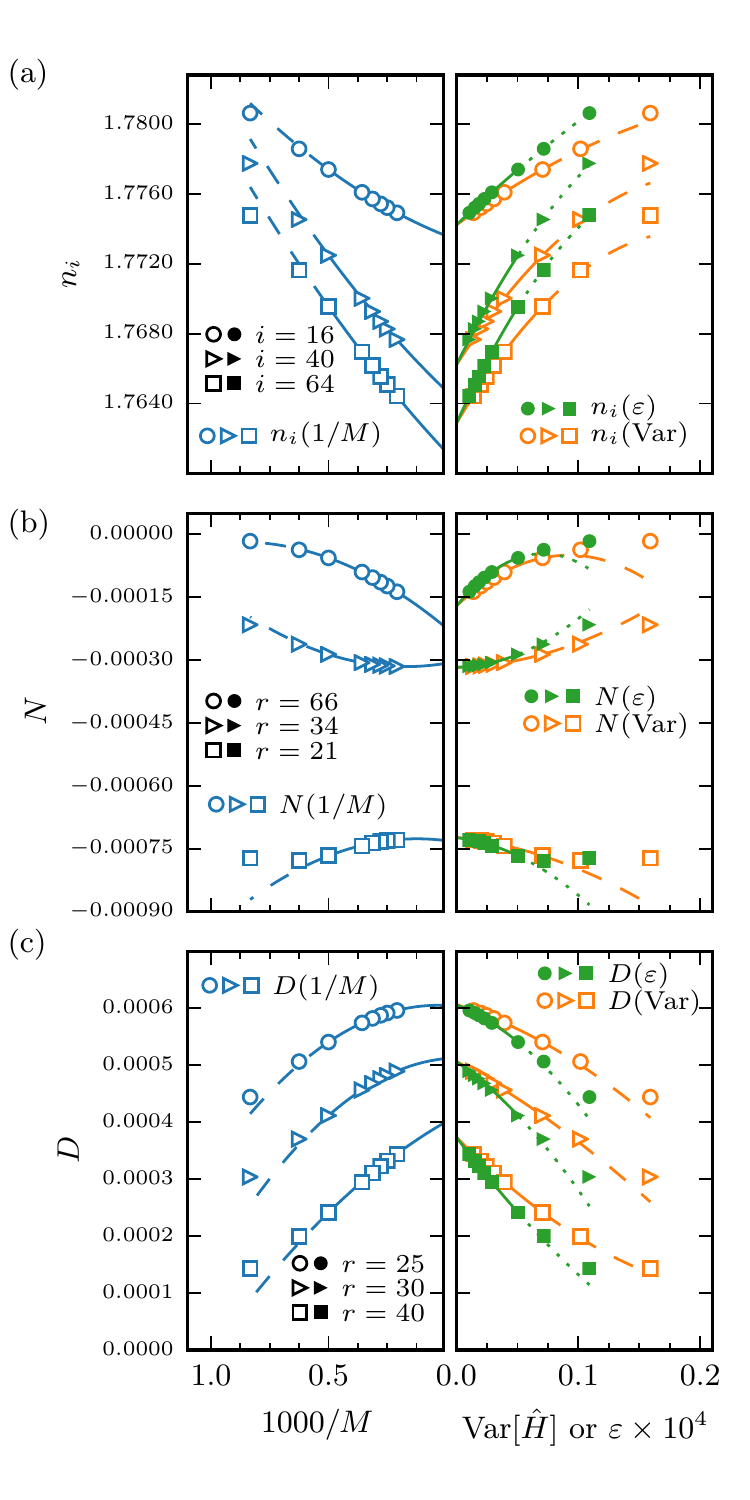}
\caption{(color online) Extrapolation of observables as a function of the inverse bond dimension $1/M$ (left panels), the truncated weight $\varepsilon$ (right panels, full symbols) and as a function of the energy variance $\Var{\hat H}$ (right panels, empty symbols). Shown are representative examples of (a) the local rung density $n_i$, (b) the density correlation function $N(r)$, and (c) the pair correlation function $D(r)$. Results are for $L=128$ and $n=0.875$.}
\label{fig:extrapolations}
\end{figure}

This strategy cannot be generalized to generic observables that do not commute with the Hamiltonian. We thus perform linear regressions with quadratic polynomials in both the energy variance, the truncated weight and $1/M$, using results for the six largest values of $M$. Spatially dependent quantities, such as the local density and the correlation functions, are independently extrapolated at each point. As a consistency test sum rules are checked, {\it e.g.} the sum of all local densities should give the total particle number.

Figure~\ref{fig:extrapolations} shows the extrapolations for $n_i$, $N(r)$ and $D(r)$ at various positions using both extrapolation schemes. One notices at first that the two approaches agree reasonably with the disagreement limited to a few percent. For local observables, such as the local density in panel (a), the difference can be very small, and deviations are generally found to be smaller towards the edge of the system. Correlation functions like $N(r)$ and $D(r)$ are harder to extrapolate as $r$ increases and the relative differences in extrapolated values grow. Accurately describing the behavior of correlation functions at longer distances requires an increasing bond dimension $M$. Higher order terms in the extrapolation thus become more important at constant $M$ and extrapolations are harder. 

\section{\label{sec:friedel-oscillations}Determining $K_\rho$ from density oscillations}

The most reliable estimation of the correlation exponent $K_\rho$ in DMRG calculations is based on density oscillations (Friedel oscillations) induced by the boundaries of the open systems commonly studied in DMRG.\cite{white2002} This method has been successfully applied to $t$-$J$ ladders,\cite{white2002} quantum wires\cite{weiss2007} and supersymmetric fermions.\cite{bauer2013} 

Friedel oscillations observed in the local density profile take the form
\begin{equation}
\label{eq:friedel-oscillations}
n(x) \equiv \langle \hat n(x) \rangle \approx A \frac{\cos(2 \pi N_h x / L_{\mathrm{eff}} + \phi_1)}{\left[ L_{\mathrm{eff}} \sin(\pi x / L_{\mathrm{eff}} + \phi_2) \right]^{K_\rho/2}} + n_0
\end{equation}
where $A$ is a non-universal amplitude, $\phi_1$ and $\phi_2$ phase shifts, $n_0$ the background density, $N_h$ is the number of holes in the system and $ L_{\mathrm{eff}}\sim L$ is an effective length. To derive this expression, Ref. \onlinecite{white2002} considered the slowest decaying component of the density-density correlation function in the Luther-Emery model, finite size effects are then introduced via a standard conformal transformation.

\begin{figure}[t!]
\centering
\includegraphics[width=\columnwidth]{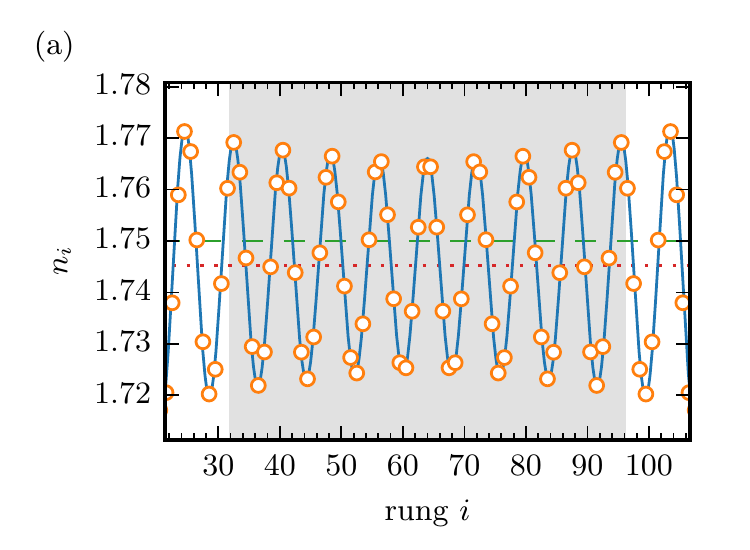} \vspace{-3em} \\
\includegraphics[width=\columnwidth]{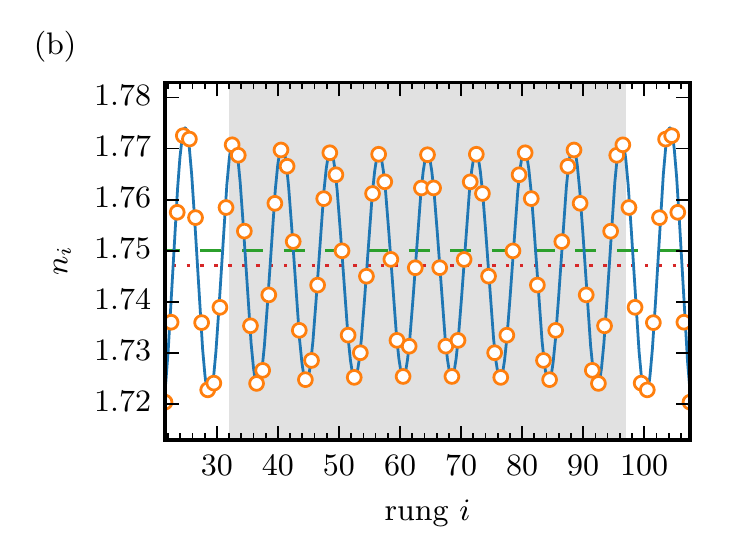} \vspace{-3em} \\
\includegraphics[width=\columnwidth]{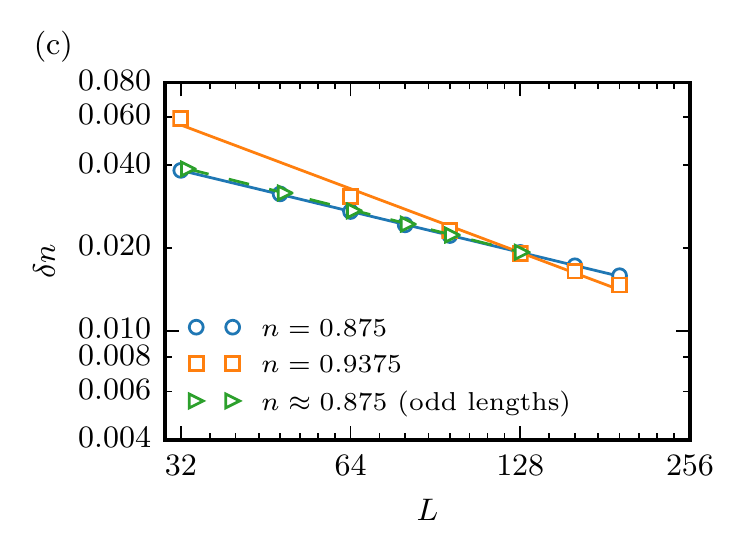}
\caption{(color online) (a) Fit of the density profile (solid line) for a system of length $L=128$ with filling $n=0.875$ compared to the raw density (markers). The red dotted line is the density offset $n_0$ obtained from the least-square fit, as a reference, the dashed green line shows the average filling $n$. The fit is restricted to the shaded region to avoid the divergent boundaries. (b) Same as (a) for a system with an additional rung but the same number of holes, i.e. with $N_\uparrow=N_\downarrow=113$ (two more particles and the same number of holes). (c) Finite size scaling of the oscillation amplitude $\delta n(L)$ as a function of the system size $L$ in double logarithmic plot for several filling parameters.}
\label{fig:density-fit}
\end{figure}

The need for an effective length can be understood as an effect of the finite extent of the hole pairs in the ladder. We find that an effective length $L_{\mathrm{eff}} = L - 2$ best describes our results. For our choice of doping the ladders with multiples of four holes, there is a density maximum in the center of the ladder and we thus need to fix the phase shifts to $\phi_1 = -\pi N_h L / L_{\mathrm{eff}}$ and $\phi_2=\frac{\pi}{2} (1 -  L / L_{\mathrm{eff}})$.

The exponent $K_\rho$ can be obtained from finite size scaling at constant density $\rho$. In particular, the density oscillation in the center of the system follow
\begin{equation}
\delta n(L)=n(L/2)-n_0 \sim L_\mathrm{eff}^{-K_\rho/2}.
\end{equation}

Unfortunately, extracting the oscillation amplitude is not as straight-forward. The first problem is that the background density $n_0$ is not simply the mean density $n$, but instead depends on $L$. This can be seen by integrating Eq.~\eqref{eq:friedel-oscillations} and equating it to the total particle number. Numerically, this deviation can be observed in the top panel of Fig.~\ref{fig:density-fit}. Secondly, for an even number of rungs, the finite lattice spacing limits the spatial resolution, hence often it is not possible to obtain the density exactly at $x=L/2$, where the oscillating factors are trivial. This is not the case for an odd number of rungs as one can see in Fig.~\ref{fig:density-fit}b. From the lower panel of Fig.~\ref{fig:density-fit} we note that the latter problem has a minor impact on the final result, therefore we continue the analysis with ladders of even length.

To avoid the first issue we perform a non-linear least square fit of the local densities in the middle of our system to Eq.~\eqref{eq:friedel-oscillations}. The parameters $L_{\mathrm{eff}}$, $\phi_1$ and $\phi_2$ are fixed as discussed above and $A$, $K_\rho$ and $n_0$ are used as fit parameters. The obtained fit is then used to compute $\delta n(L)$, and $K_\rho$ is extracted in another fit to $\delta n(L) \sim L_\mathrm{eff}^{-K_\rho/2}$. This approach works very well, as one can see from the illustrative examples in Fig.~\ref{fig:density-fit}. Here, we also show the difference between the size-dependent background density $n _0$ and the mean density $n$. 

Other approaches for extracting the oscillation amplitude provided no reliable results. One of the failed attempts was to fix $n_0$ from a linear interpolation of the two closest points to the nodes in the oscillations. We also tried obtaining $n(L/2)$, by accounting for the $\sin$ and $\cos$ terms in the finite size scaling, but this approach was unstable because of numerical errors in obtaining the wavelength of the oscillations.

\begin{table}[t]
\centering
\caption{$K_\rho$ as extracted from the fit in Fig.~\ref{fig:density-fit}.}
\label{tab:Krho-Friedel}
\begin{ruledtabular}
\begin{tabular}{lcccccc}
& \multicolumn{2}{c}{$\Var{\hat H}$} & \multicolumn{2}{c}{$\varepsilon$} & \multicolumn{2}{c}{$1/M$} \\
$n$  & $K_\rho$ & { ${R^2}$} & $K_\rho$ & { ${R^2}$} & $K_\rho$ & { ${R^2}$} \\ \hline
0.875 & 0.99 & { ${0.9999}$} & 0.92 & { ${0.9999}$} & 1.17 & { ${0.9956}$} \\
0.9375 & 1.54 & { ${ 0.9919}$} & 1.54 & { ${0.9916}$} & 1.66 & { ${0.9973}$} \\
\end{tabular}
\end{ruledtabular}
\end{table}

Our results for the exponent $K_\rho$ obtained from the fit in Fig.~\ref{fig:density-fit}c) are summarized in Tab.~\ref{tab:Krho-Friedel}. We see that $K_\rho$ increases with filling and it is consistent with reaching the limit $K_\rho = 2$ for $n=1$. The goodness of the linear regression $R^2$ is reported to  be always larger than $99\%$, which supports the expected decay of the oscillations. Results for more dilute systems are discussed in Appendix~\ref{app:weak doping}.

\begin{figure}[t!]
\centering
\includegraphics[width=\columnwidth]{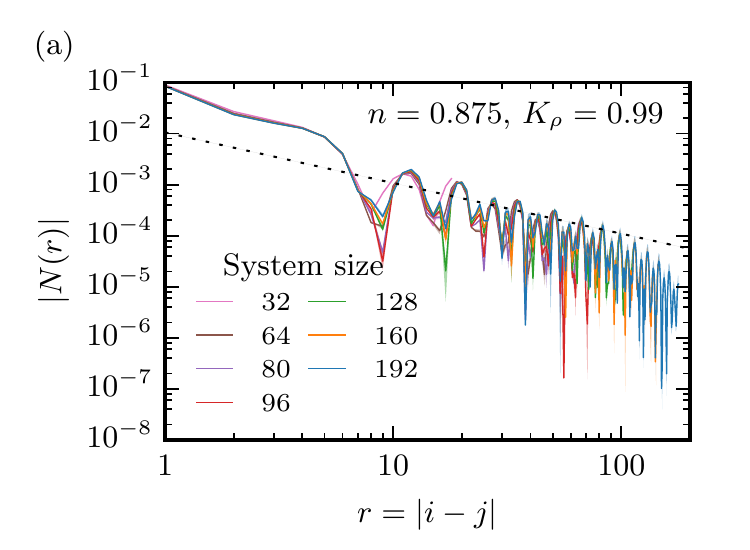} \vspace{-.5em} \\
\includegraphics[width=\columnwidth]{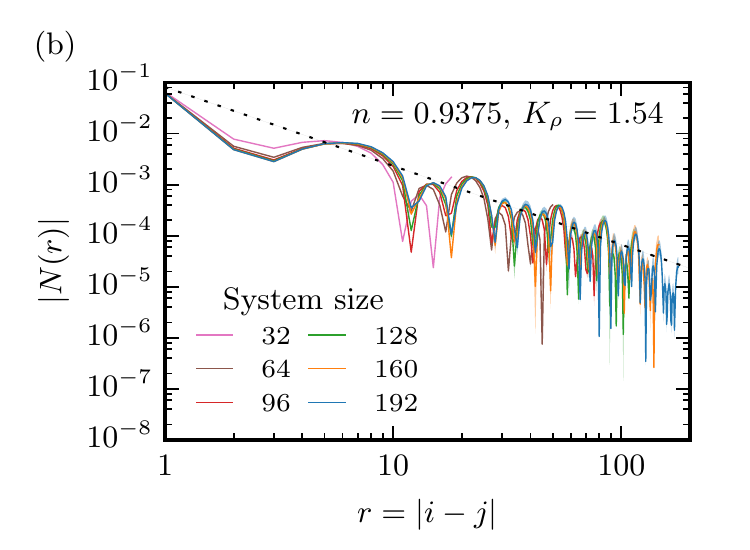}
\caption{(color online) Decay of the density-density correlation function $N(r)$ for many system system sizes. The dashed line is a power-law decay obtained with an exponent $-K_\rho$ and $K_\rho=1.54$ from the $\Var{\hat H}$ and $\varepsilon$ extrapolations in Table~\ref{tab:Krho-Friedel}; the vertical offset is chosen to show the agreement with the long-distance behavior of the correlation function. Panels a) and b) refer to two different average fillings $n=0.875$ and $n=0.9375$, respectively.}
\label{fig:density-correlation}
\end{figure}

\begin{figure}[t!]
\centering
\includegraphics[width=\columnwidth]{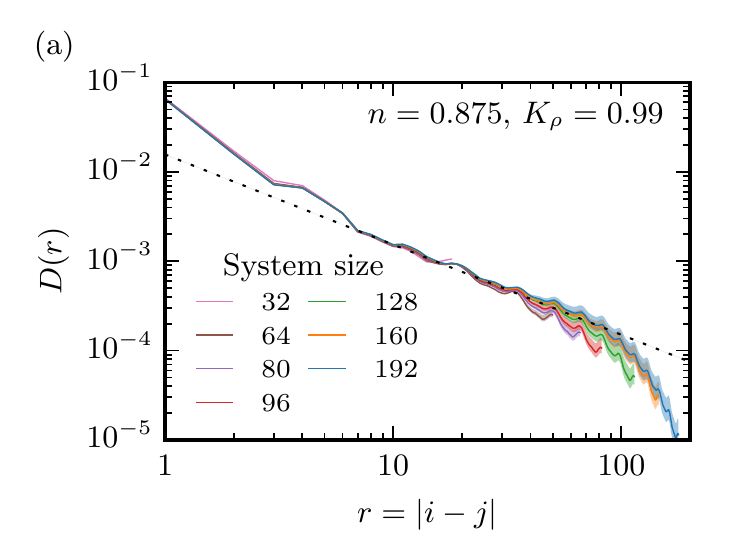} \vspace{-.5em} \\
\includegraphics[width=\columnwidth]{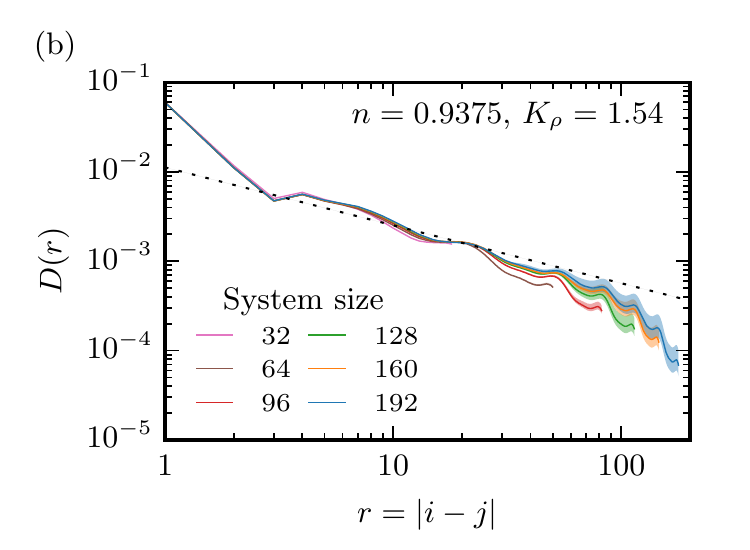}
\caption{(color online) Decay of the pair correlation function $D(r)$ for many system system sizes. The dashed line is a power-law decay obtained with an exponent $-1/K_\rho$ and $K_\rho=1.54$ from the $\Var{\hat H}$ and $\varepsilon$ extrapolations in Table~\ref{tab:Krho-Friedel}; the vertical offset is chosen to show the agreement with the long-distance behavior of the correlation function. Panels a) and b) refer to two different average fillings $n=0.875$ and $n=0.9375$, respectively.}
\label{fig:pairfield}
\end{figure}

\section{\label{sec:correlations}Correlation functions}

Our results at low doping are consistent with the expected dominant superconducting correlations. The correlation exponent obeys $K_\rho > 1$, and therefore superconducting pair correlations dominate and decay slower than $1/r$. This is in contrast to Fig. 4 of Ref.~\onlinecite{noack1997}, where the pair correlations decay roughly as $r^{-1.5}$ at $n=0.875$ and $n=0.9375$. 

This puzzling discrepancy can be resolved by noting that it is hard to faithfully obtain the correlation exponent from the spatial decay of correlation functions. In particular, as we argued in Sec.~\ref{sec:method}, large system sizes $L$ and large bond dimensions $M$ are needed to obtain reliable results for correlation functions. Both finite $M$ and finite $L$ suppress long range correlations. While the system sizes and bond dimensions of Ref.~\onlinecite{noack1997} are sufficient to converge local quantities, they are insufficient for correlation functions, and in particular the pair correlations.

Since the determination of $K_\rho$ from local densities via Friedel oscillations is expected to be more reliable than from the decay of the correlation functions, we here use the values obtained in the previous section and show that the correlation functions -- after proper extrapolation in $M$ and $L$ -- are consistent with these values. 
Figures ~\ref{fig:density-correlation} and~\ref{fig:pairfield} show $N(r)$ and $D(r)$ for various system sizes. We notice three regimes: At short distances we find a non-universal regime of fast-decaying correlation functions. At the longest distances finite-size effects become relevant and the correlation functions are strongly suppressed. In between we find a region where the spatial decay of the correlation functions is indeed consistent with a power law. For $L=192$ and extrapolating $M\to \infty$ we find good agreement in the range $10 \lesssim r \lesssim 90$ with the expected behavior based on the values of $K_\rho$ obtained by the extrapolations in $\Var{\hat H}$ and $\varepsilon$ in Tab.~\ref{tab:Krho-Friedel}.

Note that where the pairfield correlations always remain positive, the density correlations oscillate across zero. This eventually leads to the spikes when $N(r)$ is about to change sign in Fig.~\ref{fig:density-correlation}, where we plot its absolute value on a double-logarithmic axis.

\section{\label{sec:conclusions}Conclusions}
In this paper we settle the long-standing disagreement between the analytically predicted behavior of the pair correlation functions in weakly doped Hubbard ladders and results of DMRG calculations. We illustrate the two main causes of the discrepancy in previous results, which had indicated that the pair correlation function decays faster than expected. The first cause is the need for very long system sizes, as finite sizes tend to strongly suppress pair correlation functions when the distance becomes comparable to the system size. More importantly, a careful extrapolation in the bond dimension to $M\to \infty$ is necessary and has to be performed separately for each distance. Increasingly larger bond dimensions $M$ are needed to obtain converged results for longer distances $r$, substantially larger than necessary to converge local quantities.

We devote particular attention to the extrapolation techniques. All data used for our work and fitting and extrapolation workflows are available as Supplementary Material~\cite{supplementary} to allow the reader to reproduce our results and modify the details of the extrapolation and fit approaches and see how they affect the final results.

While we here use a standard finite-size DMRG approach with open boundary conditions, recently proposed techniques such as the sine-square deformation~\cite{hikihara2011}, grand-canonical DMRG~\cite{hotta2012} or infinite-size DMRG~\cite{mcculloch2008} could also be applied to this problem.

Achieving a good understanding of the two-leg ladder and the effects of finite entanglement scaling and finite size scaling of DMRG observables, and obtaining reliable results for the pair correlation function in this simple model is an important milestone to improving the reliability of numerical simulations for larger, two-dimensional systems.

\begin{acknowledgments}
We thank Jan Gukelberger, Adrian Kantian and Mauro Iazzi for helpful discussions, and we are especially thankful to Steven R. White for pointing out to us the use of an effective length in the analysis of Friedel oscillations.
The simulations were performed using the ALPS MPS code~\cite{dolfi2014,bauer2011-alps,albuquerque2007} on the M\"onch cluster of ETH Zurich and on supercomputers of the Swiss National Supercomputing Centre (CSCS). SK acknowledges support by ETH Research Grant ETH-34 12-2. MT acknowledges hospitality of the Aspen Center for Physics, supported by NSF grant PHY-1066293.
\end{acknowledgments}

\appendix

\section{\label{app:weak doping}Weak hole-doping results}

For very dilute systems such as $n=0.96875$ we find that the finite size analysis described in Section~\ref{sec:friedel-oscillations} becomes less reliable, as DMRG convergence and extrapolation become more challenging. Since we evaluate an even number of hole pairs, only three system sizes are available, $L=64$, $L=128$ and $L=192$, corresponding to only 2, 4 and 6 hole pairs, respectively. Longer system sizes would be needed to perform a rigorous scaling analysis. Furthermore, distributing six hole pairs in such a long system is a very slow process and often leads to convergence problems as has been observed for the $t$-$J$ model.~\cite{white2015} To improve convergence speed one could employ multigrid techniques.~\cite{dolfi2012}

\begin{figure}[b!]
\centering
\includegraphics[width=\columnwidth]{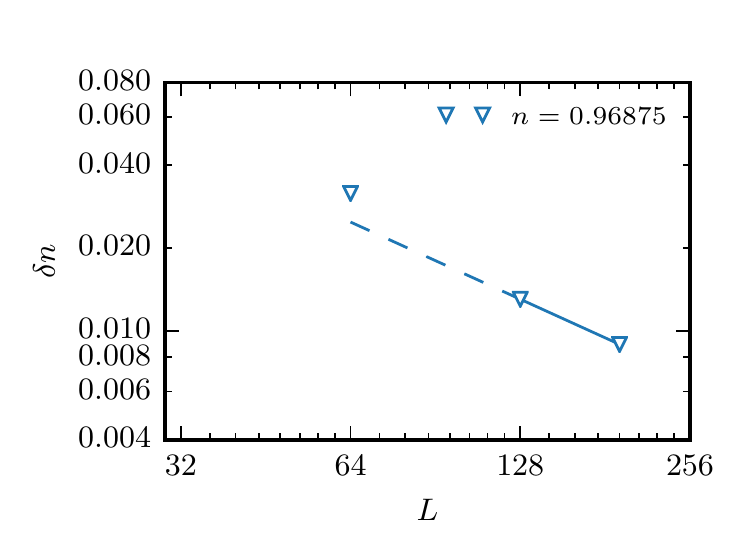}
\caption{(color online) Finite size scaling of the oscillation amptitude $\delta n(L)$ as a function of the system size $L$ in double logarithmic plot for several an average filling $n=0.96875$.}
\label{fig:density-fit-weakdoping}
\end{figure}

\begin{figure}[p!]
\centering
\includegraphics[width=\columnwidth]{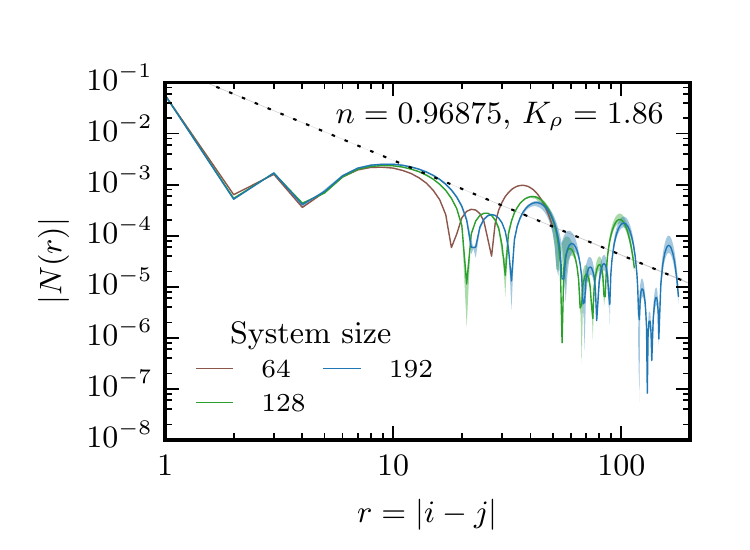}
\caption{(color online) Decay of the density-density correlation function $N(r)$ for many system system sizes with average filling $n=0.96875$. The dashed line is a power-law decay obtained with an exponent $\mu=-K_\rho$ and $K_\rho$ from the $\Var{\hat H}$ and $\varepsilon$ extrapolations in Table~\ref{tab:Krho-Friedel}; the vertical offset is chosen to show the agreement with the long-distance behavior of the correlation function.}
\label{fig:density-correlation-weakdoping}
\end{figure}

\begin{figure}[p!]
\centering
\includegraphics[width=\columnwidth]{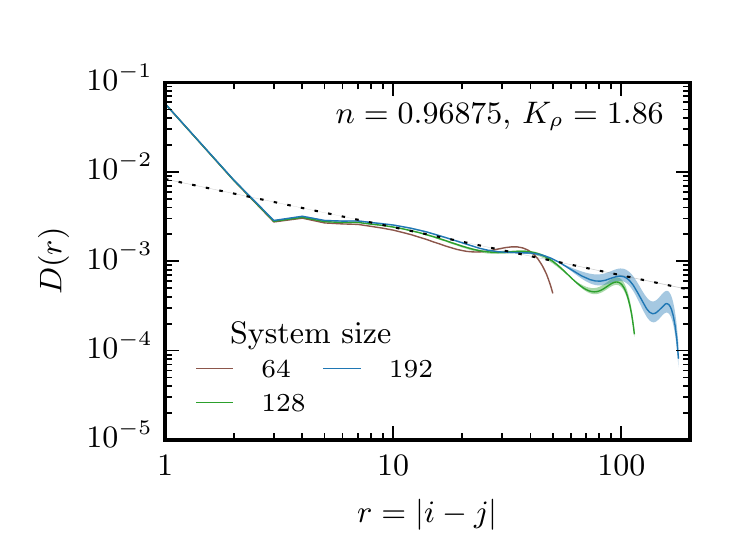}
\caption{(color online) Decay of the pair correlation function $D(r)$ for many system system sizes with average filling $n=0.96875$. The dashed line is a power-law decay obtained with an exponent $\nu=-1/K_\rho$ and $K_\rho$ from the $\Var{\hat H}$ and $\varepsilon$ extrapolations in Table~\ref{tab:Krho-Friedel}; the vertical offset is chosen to show the agreement with the long-distance behavior of the correlation function.}
\label{fig:pairfield-weakdoping}
\end{figure}

Here we present the results obtained at average filling $n=0.96875$. Figure~\ref{fig:density-fit-weakdoping} shows that finite size scaling analysis, whose exponents are reported in Table~\ref{tab:Krho-Friedel-weakdoping}. Comparisons for the spatial decay of correlation functions with the exponent $K_\rho$ expected from the Friedel oscillations are shown in Figure~\ref{fig:density-correlation-weakdoping} and~\ref{fig:pairfield-weakdoping}.

\begin{table}[t!]
\centering
\caption{$K_\rho$ as extracted from the fit in Fig.~\ref{fig:density-fit-weakdoping}, where only the two largest system sizes $L=128$ and $L=192$ are considered.}
\label{tab:Krho-Friedel-weakdoping}
\begin{ruledtabular}
\begin{tabular}{lccc}
$n$& $\Var{\hat H}$ & $\varepsilon$ & $1/M$ \\ \hline
0.96875 & 1.87 & 1.85 & 2.39
\end{tabular}
\end{ruledtabular}
\end{table}

Note that in the scaling analisys of the Friedel oscillations we consider only the two largest system sizes $L=128$ and $L=192$, because the first system size $L=64$ contains only two hole pairs, hence finite size effects are expected to have a dominant contribution. The value of $K_\rho$ obtained from the fit (see Table~\ref{tab:Krho-Friedel-weakdoping}) is again compatible with the expected limit for $n=1$, and it is compatible with the decay of correlation functions (see Figure~\ref{fig:density-correlation-weakdoping} and~\ref{fig:pairfield-weakdoping}).

\bibliographystyle{apsrev4-1}
\bibliography{refs}

\end{document}